\documentclass[preprint,12pt]{elsarticle}

\usepackage{graphicx}
\usepackage[section]{placeins} 
\usepackage{amsmath}
\usepackage{color}
\usepackage{rotating}

\usepackage{lineno,hyperref}
\usepackage{amsmath}
\usepackage{adjustbox}
\usepackage{tikz}
\usepackage{graphicx}
\usepackage{float}
\usetikzlibrary{positioning}
\usetikzlibrary{shapes, arrows}
\usetikzlibrary{shapes}
\usepackage{mathtools}
\usepackage{csquotes}
\usepackage{amssymb}
\usepackage{mathrsfs}
\usepackage{mathtools}
\usepackage{xcolor}
\usepackage{amssymb}
\usepackage{nameref}
\usepackage[toc,page]{appendix}
\usepackage[english]{babel}
\usepackage[T1]{fontenc}
\usepackage{lipsum}
\usepackage{caption}
\usepackage{multirow}
\usepackage{array}
\usepackage{booktabs}
\usepackage{relsize}
\usepackage[percent]{overpic}
\usepackage{multirow}

\usepackage{amssymb}
\usepackage{amsmath}

\newcommand{\sz}[1]{{{\color{black}#1}}}

\newcommand{\nsc}[1]{{\color{black}#1}}
\newcommand{\agr}[1]{{\color{black}#1}}
\begin{document}

\begin{frontmatter}



\title{Evaporation of Finite-Size Ammonia and n-Heptane Droplets in Weakly Compressible Turbulence: An Interface-Resolved DNS Study}

\author[NTNU]{Salar Zamani Salimi\corref{mycorrespondingauthor}}
\cortext[mycorrespondingauthor]{Corresponding author: salar.z.salimi@ntnu.no}
\author[NTNU,Sintef]{Andrea Gruber}
\author[Princeton]{Nicolò Scapin}
\author[NTNU,TURIN]{Luca Brandt}
%
%
\address[NTNU]{Department of Energy and Process Engineering, Norwegian University of Science and Technology (NTNU), Trondheim, Norway}
\address[Sintef]{SINTEF Energy Research, Thermal Energy Department, Trondheim, Norway}
\address[Princeton]{Department of Mechanical and Aerospace Engineering, Princeton University, New Jersey, USA}
\address[TURIN]{Department of Environment, Land and Infrastructure Engineering (DIATI), Politecnico di Torino, Torino, Italy}

\begin{abstract}
This study presents direct numerical simulation (DNS) of finite-size, interface-resolved ammonia and n-heptane droplets evaporating in decaying homogeneous isotropic turbulence. Simulations are conducted for each fuel to model the dynamics in a dense spray region, where the liquid volume fraction exceeds $\mathcal{O}(10^{-2})$. The focus is on investigating the complex interactions between droplets, turbulence, and phase change, with emphasis on droplet-droplet interactions and their influence on the evaporation process. 
The present study also explores how varying turbulence intensities affect the evaporation rates of each fuel, unveiling the differences in the coalescence and energy transfer from the liquid to the gaseous phase. The results reveal that, when comparing ammonia with n-heptane with equal liquid volume fractions, ammonia exhibits faster initial evaporation due to its higher volatility. However, this rate declines over time as frequent droplet coalescence reduces the total surface area available for evaporation. When numerical experiments are initialized with equal energy content, increasing turbulence intensity enhances the evaporation of n-heptane throughout the simulation, while ammonia evaporation soon becomes less sensitive to turbulence due to rapid vapor saturation. These findings are relevant to improving predictive CFD models and optimizing fuel injection in spray-combustion applications, especially under high-pressure conditions.

\end{abstract}

\begin{keyword}
Droplets, Turbulence, Evaporation, Spray combustion, Direct numerical simulation, Volume of fluid
\end{keyword}
\end{frontmatter}



\section{Introduction}

Gas-liquid turbulent flows involving phase change play a key role in many natural and engineering processes, such as rain formation, spray cooling, and spray combustion in propulsion systems~\cite{BIROUK2006408,JENNY2012846}. \nsc{For instance}, in high-pressure combustion engines, the intricate interactions between \agr{liquid-fuel atomization, droplets} evaporation, combustion, and turbulence significantly influence both \agr{ignition characteristics, combustion} efficiency and emissions \agr{of regulated pollutants}. To create accurate predictive models and enhance the design of combustion devices, a \agr{more complete} understanding of the complex interplay between turbulence and evaporation in a droplet cloud under high-pressure conditions is essential. \par
In spray-combustion processes, accurately describing droplet interactions requires taking into account the liquid volume fraction, $\alpha_l$, which has been shown to be $\mathcal{O}(10^{-3})$ near the flame zone~\cite{WANG20193335}, and higher closer to the injector. Generally, spherical droplets in homogeneous isotropic turbulence (HIT) can be classified by their diameter $d_0$ relative to the smallest turbulent length scale (the Kolmogorov scale, $\eta$) as sub-Kolmogorov size ($d_0 \ll \eta$) or finite size ($d_0 > \eta$). For subkolmogorov size, droplets are usually treated as point particles, and point-particle methods have been widely used in numerical simulations of droplet-laden flows with phase change~\cite{KUERTEN2016}. In these approaches, the governing equations for the liquid phase are simplified to a set of Lagrangian equations, which are solved for each droplet individually. These droplets' dynamics are then coupled with the gas-phase Eulerian equations \agr{using one- or two-way inter-phase momentum exchange~\cite{GUALTIERI2015} and semi-empirically formulations to evaluate the evaporative fluxes~\cite{MILLER19981025}.} \par
Direct numerical simulations (DNS) of evaporating droplets interacting with turbulence have been performed using the point-particle method in homogeneous shear flows~\cite{MASHAYEK1998163}. Additionally, DNS with point-particles has been widely applied to investigate droplets in spray-combustion systems and reactive flows~\cite{REVEILLON20072319,Xia20112581,LUO20112143,VIE20151675}. \par
Recent advances in phase-change modeling and numerical algorithms \cite{TANGUY2007837,Welch2000662,SCHLOTTKE20085215,CIPRIANO2024112955,SCAPIN2020,ZAMANISALIMI2024,SALARJCP2025} have made it possible to simulate turbulent flows around evaporating interface-resolved droplets with sizes comparable to or larger than the Kolmogorov scale ($d_0 / \eta \geq 1$), known as "finite-size" droplets~\cite{scapin_lupo,DODD2021}. While, most of these studies have focused on the dilute spray regions, where the liquid volume fraction is low, they have nevertheless demonstrated the potential of interface-resolved numerical simulations. From a modelling perspective, we have learned that, existing models based on empirical laws are accurate enough at high temperatures and low concentrations, while they lose accuracy at lower pressure and temperatures. \par
In the present work, we present results from  DNS of finite-size deformable droplets in decaying homogeneous isotropic turbulence, focusing on the dense spray region where the initial liquid volume fraction exceeds $\mathcal{O}(10^{-2})$. In this regime, not only resolving the exchange of mass, momentum, and heat between the multi-component gas phase and the liquid droplets remains crucial, but accurately capturing droplet-droplet interactions becomes equally important. Furthermore, we investigate the evaporation of two different liquid fuels, ammonia, and n-heptane, to provide state-of-the-art comparative insights about their evaporation characteristics. \agr{N-heptane is a widely used surrogate for light diesel—commonly employed in propulsion and distributed power-generation applications involving combustion engines whereas ammonia turbines have recently emerged as a promising carbon-free energy vector for similar uses, owing to its favorable transport and storage logistics~\cite{VALERAMEDINA2018}. Nevertheless, challenges persist that hinder the development of clean and efficient ammonia-fired combustion systems, primarily due to a lack of detailed experimental characterization and a fundamental understanding of the critical differences in atomization, evaporation, and ignition processes. Therefore, the current study aims to provide a more comprehensive comparative analysis of the evaporation processes characterizing these two distinct fuels.}.

\section{Mathematical formulation}
\nsc{In this section, we discuss the governing equations and the numerical methodology utilized for the current simulations of droplet evaporation in homogeneous isotropic turbulence in the low-Mach limit.}
\subsection{Governing equations}
We consider a two-phase (gas-liquid) turbulent flow with phase change in the low Mach-number limit. In this regime, the liquid phase is treated as incompressible with constant properties, while the gas phase is compressible, allowing its properties to vary with temperature, thermodynamic pressure, and composition. This weakly compressible formulation captures significant density variations in the gas phase bulk region while filtering out acoustic effects. The system comprises two immiscible Newtonian fluids: a single-component liquid and a multi-component gas, which is an ideal mixture of inert gas and a vaporized liquid. The evaporation process is driven by the partial pressure of the inert gas in the gas phase. To describe this system, we define a phase indicator function, $C$, at position $\mathbf{x}$ and time $t$, to distinguish between the two phases.
\begin{equation}
C(\mathbf{x}, t) =
\begin{cases} 
      1 & \text{if } x \in \Omega_l\mathrm{,} \\
      0 & \text{if } x \in \Omega_g\mathrm{,}
\end{cases}
\end{equation}
where $\Omega_l$ and $\Omega_g$ represent the domains corresponding to the liquid and gas phases, respectively, which are separated by a zero-thickness interface, $\Gamma(t) = \Omega_l \cap \Omega_g$. Throughout this manuscript, the subscripts $l$ and $g$ are used to denote liquid-phase and gas-phase properties. The conservation laws governing momentum, vapor species, thermal energy, and mass across the interface are expressed as follows:
\begin{equation}
	\rho \frac{D\textbf{u}}{Dt}= -\boldsymbol{\nabla} p + \frac{1}{Re}\boldsymbol{\nabla} \cdot \boldsymbol{\tau} + \frac{\mathcal{F}_\sigma}{We}\mathrm{,}
	\label{eq1}
\end{equation}
\begin{equation}
	\rho_{g} \frac{D \omega}{Dt} = \frac{1}{ReSc}\boldsymbol{\nabla} \cdot (\rho_g \mathcal{D}\boldsymbol{\nabla} \omega)\mathrm{,}
	\label{eq2}
\end{equation}
\begin{equation}
\begin{split}
		&\rho c_p \frac{DT}{Dt}=\frac{1}{RePr}\boldsymbol{\nabla} \cdot (k \boldsymbol{\nabla} T) + \Biggl(\Theta \frac{dp_{th}}{dt}+\\
         	&\frac{\rho_g \mathcal{D}}{ReSc}\sum_{j=1}^{2}\boldsymbol{\nabla} h_j\cdot\boldsymbol{\nabla} \omega_j \Biggr)(1-C)-\frac{(\dot{m}_\Gamma \delta_\Gamma)}{Ste},
	\label{eq3}
\end{split}
\end{equation}
\begin{equation}\label{eq:4}
	\dot{m}_{\Gamma} = \frac{1}{ReSc}\frac{\rho_{g,\Gamma}\mathcal{D}_{\Gamma}}{1-\omega}\boldsymbol{\nabla}_\Gamma \omega \cdot \textbf{n}_\Gamma.                 
\end{equation}
In this context, $\mathbf{u}$ denotes the fluid velocity, which is assumed to remain continuous across both phases except for the interface where phase change occurs, $p$ represents the hydrodynamic pressure, and $T$ is the temperature. The enthalpy, $h$, is expressed as $h = c_p T$ with its gradient, $\boldsymbol{\nabla} h = \boldsymbol{\nabla} (c_p T)$. The mass fraction of vaporized liquid in the inert gas is represented by $\omega$, and $\dot{m}_\Gamma$ denotes the interfacial mass flux. In equation (\ref{eq1}), $\boldsymbol{\tau}$ refers to the viscous stress tensor for compressible Newtonian flows, and $\mathcal{F}_\sigma = \kappa_{\Gamma} \delta_\Gamma$, where $\kappa_\Gamma$ is the interfacial curvature. The general thermophysical property, $\zeta$ (which includes density $\rho$, dynamic viscosity $\mu$, thermal conductivity $k$, or specific heat capacity $c_p$), is calculated as an arithmetic average: 
\[ \zeta = 1 + (\lambda_{\zeta} - 1)C, \quad \textrm{with} \quad \lambda_{\zeta} = \zeta_l / \zeta_{g,r}.\] 
Since the liquid property $\zeta_l$ is constant and uniform, further modeling is unnecessary. The gas-phase properties $\zeta_g$ are determined using appropriate equations of state. The gas density, $\rho_g$, is obtained from the ideal gas law, and the vaporized liquid diffusion coefficient, $\mathcal{D}$, is calculated using the Wilke-Lee correlation~\cite{Reid}. Additional details on the way the gas thermophysical properties are evaluated can be found in~\cite{scapin_lupo}.
\par 
In  equations \ref{eq1}-\ref{eq:4}, different dimensionless parameters appear. By introducing a reference velocity $u_r$ and reference length $l_r$, we define the Reynolds number $Re = \rho_{g,r} u_r l_r / \mu_{g,r}$, the Weber number $We = \rho_{g,r} u_r^2 l_r/\sigma$, with $\sigma$ the surface tension; $Sc = \mu_{g,r}/(\rho_{g,r}\mathcal{D}_{r})$ and $Pr = \mu_{g,r} c_{pg,r}/k_{g,r}$ are the Schmidt and Prandtl numbers. Note
that the temperature equation~\ref{eq3} requires the definition of the Stefan number $Ste = c_{pg,r} T_{g,0}/\Delta h_{lv}$, where $T_{g,0}$ is the initial gas temperature and $\Delta h_{lv}$ is the latent heat, and
of the dimensionless group $ \Theta = R/(c_{pg,r} M_g )$, where $M_g$ is the molar mass of the gas
phase and $R$ the universal gas constant. \par
To close the system of equations, two additional equations are needed: one to account for the velocity divergence and another to update the thermodynamic pressure, $p_{\text{th}}$, when dealing with closed or periodic domains, ensuring mass conservation.
\begin{equation}
\boldsymbol{\nabla} \cdot \mathbf{u}  = \mathcal{F}_{\mathcal{M}} + \dfrac{1}{p_{th}}\Biggr\{ \mathcal{F}_{\omega} + \mathcal{F}_{T} - \Biggr( 1 - \dfrac{\Theta}{c_p \Bar{M}_{m,av}} \Biggl)  \dfrac{dp_{th}}{dt} \Biggl\}(1-C)\mathrm{,}
\label{eqdiv}
\end{equation}
\begin{equation}
\dfrac{1}{p_{th}} \dfrac{dp_{th}}{dt} \int_{\Omega_g} \Biggr( 1 - \dfrac{\Theta}{c_p \Bar{M}_{m,av}} \Biggl) d \Omega_g = \int_{\Omega} \Biggr\{ \mathcal{F}_{\mathcal{M}} + \frac{\mathcal{F}_{T}+\mathcal{F}_{\omega}}{p_{th}} (1-C) \Biggl\} d\Omega\mathrm{.}
\label{eqpth}    
\end{equation}
In eqs.~\ref{eqdiv} and~\ref{eqpth}, $\Bar{M}_{m,av}$ represents the mixture molar mass, calculated using a harmonic average of the gas and liquid molar masses. The terms $\mathcal{F}_{\mathcal{M}}$, $\mathcal{F}_{\omega}$, and $\mathcal{F}_{T}$ denote the different contributions to the total velocity divergence: $\mathcal{F}_{\mathcal{M}}$ accounts for the effects of phase change, $\mathcal{F}_{\omega}$ reflects the change in gas density due to vapor composition, and $\mathcal{F}_T$ captures the changes due to temperature variations.

\begin{equation}
\mathcal{F}_{M} = \dot{m}_{\Gamma} \biggr( \dfrac{1}{\rho_{g,\Gamma}} - \frac{1}{\lambda_{\rho}} \biggr) \delta_{\Gamma}\mathrm{,}
\label{eqfm}
\end{equation}

\begin{equation}
\mathcal{F}_{\omega} = \dfrac{1}{Re Sc} \dfrac{\Bar{M}_{m,av}}{\rho_g} \Biggr( \dfrac{1}{\lambda_{M}} -1 \Biggr)  \boldsymbol{\nabla} \cdot ( \rho_g \mathcal{D} \boldsymbol{\nabla} \omega )\mathrm{,}
\label{eqfvap}
\end{equation}

\begin{equation}
\mathcal{F}_T = \dfrac{1}{Re} \dfrac{\Theta}{c_p \Bar{M}_{m,av}} \Biggr[ \dfrac{1}{Pr} \boldsymbol{\nabla} \cdot (k\boldsymbol{\nabla}T) + \dfrac{\rho_g \mathcal{D}}{Sc} \sum_{j=1}^{2}  \boldsymbol{\nabla} h_j \cdot \boldsymbol{\nabla}\omega_j \Biggl]\mathrm{,}
\label{eqftmp}
\end{equation}
where $\lambda_M = M_l/M_g$ is the molar mass ratio. The complete derivation of eqs.~\ref{eqdiv} and~\ref{eqpth} is shown in~\cite{scapin_lupo} and also discussed in~\cite{SCAPIN2023}.

\subsection{Numerical method}
The governing equations are discretized on a uniform Cartesian grid with a staggered arrangement for the velocity components. Convective terms are treated using the QUICK scheme, while diffusive terms are treated with central difference schemes. For time integration, the second-order Adams-Bashforth method is employed. \par
The interface reconstruction and advection are performed entirely within an Eulerian framework, utilizing an algebraic Volume of Fluid (VoF) method, specifically the MTHINC method~\cite{II2012,Rosti2019}. The evolution of the phase indicator function is governed by the transport equation:
\begin{equation}
\dfrac{\partial C}{\partial t} + \mathbf{u}_{\Gamma} \cdot \boldsymbol{\nabla} C = 0\mathrm{,}
\label{eqvof}
\end{equation}
where $\mathbf{u}_{\Gamma}$ is the interface velocity, constructed as the sum of the extended liquid velocity $\mathbf{u}_l^{e}$ (derived from the one-fluid velocity $\mathbf{u}$) and the interfacial mass flux contribution $\dot{m}_{\Gamma} \mathbf{n}_{\Gamma}/\rho_l$. Further details are provided in~\cite{SCAPIN2020,ZAMANISALIMI2024}. \par
The transport of the vapor mass fraction, described by eq.~\ref{eq2}, is computed solely within the gas domain $\Omega_g$, under the assumption of saturation conditions at the interface $\Gamma$. Consequently, the vapor mass fraction $\omega = \omega_{\Gamma}$ at the interface is a function of the thermodynamic pressure and temperature and is imposed as a Dirichlet boundary condition. To calculate $\omega_\Gamma$, the Span-Wagner equation of state is used. Gradient evaluation of $\boldsymbol{\nabla} \omega$ is performed on an irregular grid, employing one-sided finite differences for cells adjacent to the interface and by central difference schemes for cells farther away (refer to~\cite{scapin_lupo},~\cite{SCAPIN2020} for more details). Since the temperature equation is solved over the entire domain, independent of the phase location, the liquid and gas temperatures are not directly accessible. To resolve potential artifacts such as artificial heating, particularly due to significant density differences between phases, we apply a constant extrapolation of $T$ to reconstruct the gas temperature in $\Omega_g$ and the liquid temperature in $\Omega_l$. These fields, $T_g$ and $T_l$, are defined in cells adjacent to the interface and used to update the thermophysical properties. \par 
The interfacial mass flux $\dot{m}_{\Gamma}$ is computed in the gas region by projecting the interfacial gradient along the normal direction, employing a dimension-by-dimension approach. This method ensures that $\dot{m}_{\Gamma}$ is computed only at the grid nodes in the gas region. However, as indicated by eq.~\ref{eq:4}, non-zero values of $\dot{m}_{\Gamma}$  are also needed at certain points within the liquid region. To accommodate this, $\dot{m}_{\Gamma}$ is extrapolated into a narrowband surrounding the interface, within cells where $|\boldsymbol{\nabla}C| \neq 0$. \par
The velocity divergence is calculated by discretizing the right-hand side of equation~\ref{eqdiv} node-by-node, following the method proposed in~\cite{Dalla} for weakly compressible two-phase solvers without phase change. In the low-Mach framework, the thermodynamic pressure $p_{th}$ plays a pivotal role in ensuring mass conservation of the compressible gas phase at the discrete level, as it influences the calculation of the gas density. It is important to note that mass conservation cannot be achieved solely through the advection of the phase indicator function or the pressure-correction step, which only enforces volume conservation in closed and isochoric systems. To address this, $p_{th}$ is updated at each new time step $n+1$ by integrating the gas density equation $\rho_g = (p_{th,r}M_g)(p_{th}\Bar{M}_{m,av})/(T_g R T_{g,r})$ over the gas domain $\Omega_g$, following the approach outlined in~\cite{Dalla,scapin_lupo}:
\begin{equation}
p_{th}^{n+1} = \dfrac{\mathcal{M}_g^{n+1} \dfrac{p_{th,r} M_g}{R T_{g,r}}}{\displaystyle\int_\Omega \dfrac{\Bar{M}_{m,av}^{n+1}}{T^{n+1}}(1-C^{n+1}) \, d\Omega}\mathrm{,}
\label{eqpthnp1}
\end{equation}
where $\mathcal{M}_g$ represents the total gas mass, and $p_{th,r}$ and $T_{g,r}$ are the reference thermodynamic pressure and gas temperature, equal to their respective initial values. This method ensures accurate conservation of the compressible phase at the discrete level. \par
After computing the thermodynamic pressure and velocity divergence, the momentum equation~(\ref{eq1}) is solved using a standard pressure correction method. The pressure Poisson equation is addressed through a time-splitting technique, transforming the variable Poisson equation into one with constant coefficients, as proposed in ~\cite{DODD2014}. Note that as discussed in~\cite{MOTHEAU2016430} and in~\cite{Dalla}, the pressure-splitting method for solving the variable-density Poisson equation in a low-Mach framework is effective when the temperature ratio between the two phases remains below 2–3, as in all the cases in the present study.

\section{Numerical and computational setup}
We investigate the decay of three-dimensional homogeneous isotropic turbulence (HIT) within a cubic domain, applying periodic boundary conditions in all spatial directions. Air is used as the gas phase, while the liquid phase alternates between ammonia and n-heptane in separate cases; the dissolution of air into the liquid phase is neglected for simplicity. The system is initialized with a uniform pressure of 40 bar, an initial liquid temperature of 315 K, and an initial gas temperature of 750 K --conditions typical of gas turbine combustion environments~\cite{BIROUK2006408}. Two key non-dimensional parameters are varied: the initial Reynolds number based on the Taylor length scale ($Re_{\lambda}$) and the liquid volume fraction ($\alpha_l = V_l/V$). Table \ref{tab1} outlines the flow properties chosen for their relevance to aeronautical spray combustion devices~\cite{BIROUK2006408}. It is important to note that the relative volatility between ammonia and n-heptane at the considered initial liquid temperature is approximately $136$. In particular, the ratio of vapor pressure to thermodynamic pressure is $0.408$ for ammonia and $0.003$ for n-heptane.
\par
\begin{figure}[H]
\centering
\includegraphics[width = 0.40\textwidth]{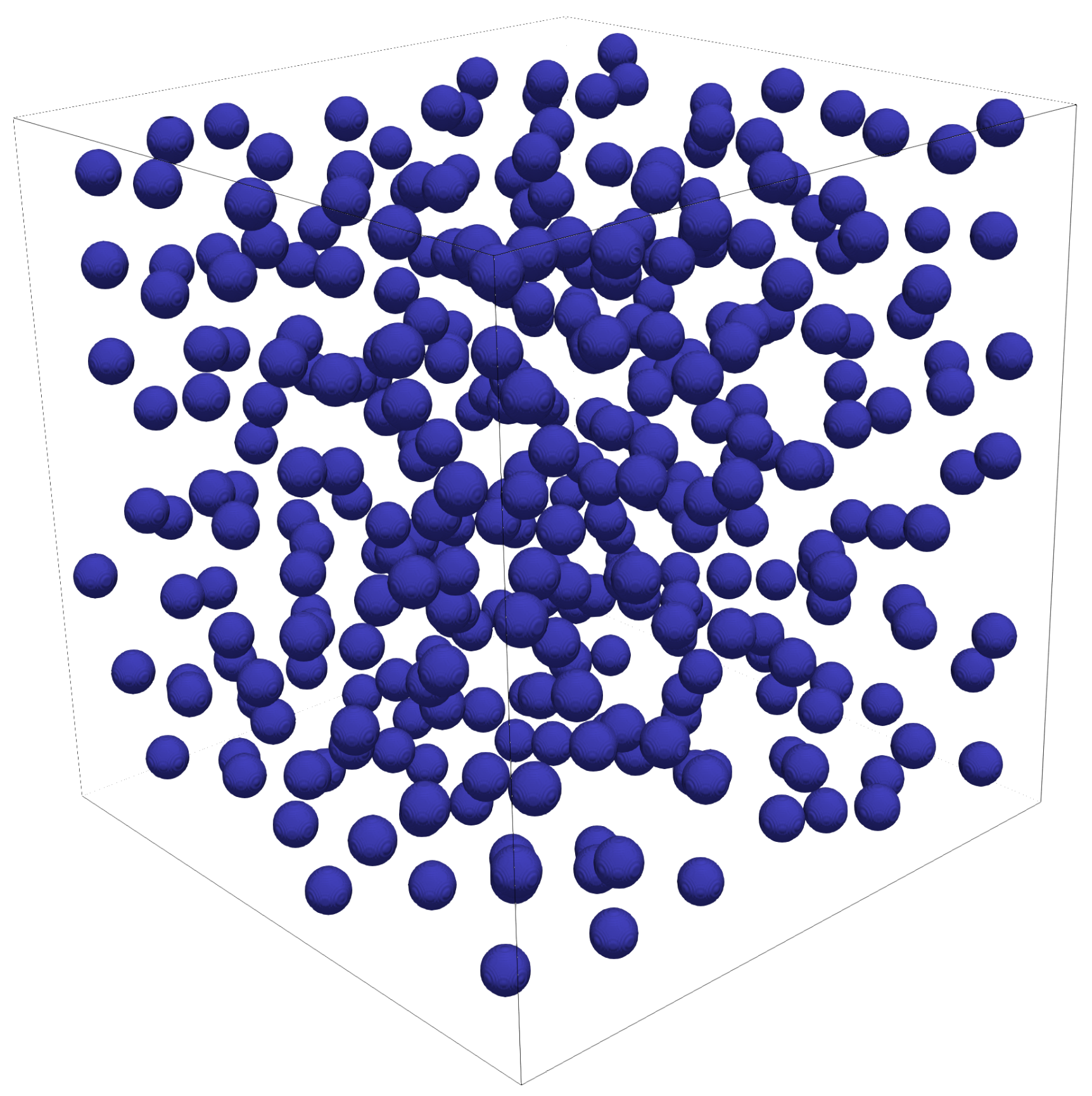}
	\caption{Initial droplet distribution: droplets are randomly seeded across the domain with a minimum center-to-center distance of $2.1d_0$ to prevent unphysical initial coalescence.} 
\label{fig:seed}
\end{figure}

For each simulation, droplets of uniform initial diameter $d_0$ are randomly seeded (Fig.~\ref{fig:seed}) into an initial velocity field that is both isotropic and divergence-free, satisfying the realizability constraint imposed by the discretized continuity equation. It is important to note that $d_0$ is equivalent for all droplets across all cases. Initially, the droplets are at rest relative to the surrounding fluid. As the simulation progresses, the turbulent kinetic energy (TKE) of the system decays as the droplets evaporate. \par
The computational domain spans $16d_0 \times 16d_0 \times 16d_0$ and is discretized using a high-resolution grid of $1024 \times 1024 \times 1024$ points. This resolution ensures the accurate representation of the flow field, with a grid spacing to Kolmogorov length scale ratio, $\Delta x/\eta \approx 0.45$ for $Re_{\lambda} \approx 80$ and $\Delta x/\eta \approx 0.91$ for $Re_{\lambda} = 140$. Each droplet is well-resolved, with an initial resolution of 64 grid points per diameter. This is consistent with our previous findings~\cite{SCAPIN2020}, which indicate that at least 50 grid points per droplet are required to fully resolve the mass, momentum, and energy exchange across the droplet interface. Furthermore, this grid resolution is sufficient to resolve the smallest scales of the two active scalar fields, vapor mass fraction ($\omega$) and temperature ($T$), as their corresponding Batchelor scales, $\eta_{B,\Omega} = \eta / \sqrt{Sc}$ and $\eta_{B,T} = \eta / \sqrt{Pr}$, exceed the Kolmogorov scale.
\begin{table}[t]
\centering
\caption{Physical parameters: type of liquid phase, initial liquid volume fraction $\alpha_l$, initial number of droplets, initial Taylor-microscale Reynolds number $Re_{\lambda} = \rho_{g,r}u_{\text{rms},r} \lambda/{\mu_{g,r}}$, initial droplet Weber number $We_{\text{rms}} = \rho_{g,r} u_{\text{rms,r}}^2 l_r/{\sigma}$, gas-phase Prandtl number $Pr = c_{pg,r} \mu_{g,r}/{k_{g,r}}$, gas-phase Schmidt number $Sc = \mu_{g,r}/{\rho_{g,r} \mathcal{D}_{r}}$, Stefan number $Ste = c_{pg,r} T_{g,r}/{\Delta h_{lv}}$, the initial gas temperature over the critical temperature $T_{g,0}/T_c$,
and $\lambda_{\Lambda}$, which is the ratio of a specific thermophysical property of the liquid phase to that of the gas phase, where $\Lambda$ could refer to density, viscosity, thermal conductivity, heat capacity, or molar mass.}

{\scriptsize
\setlength{\tabcolsep}{5pt}
\renewcommand{\arraystretch}{1.3}
\begin{tabular}{ccccccc}
\hline
Case & 1 & 2 & 3 & 4 & 5 & 6 \\
\hline
Liquid & Ammonia & n-Heptane & n-Heptane & Ammonia & n-Heptane & Ammonia \\
$\alpha_l \; (\times 10^{-2})$ & $2.1$ & $2.1$ & $1.5$ & $4.2$ & $1.5$ & $4.2$ \\
$N_d$ & $164$ & $164$ & $117$ & $328$ & $117$ & $328$ \\
$d_0/\eta$ & $29.16$ & $29.16$ & $29.16$ & $29.16$ & $58.73$ & $58.73$ \\
$Re_{\lambda}$ & $80$ & $80$ & $80$ & $80$ & $140$ & $140$ \\
$We_{rms}$ & $0.1$ & $0.1$ & $0.1$ & $0.1$ & $1.0$ & $1.0$ \\
$Pr$ & $0.71$ & $0.71$ & $0.71$ & $0.71$ & $0.71$ & $0.71$ \\
$Sc$ & $0.87$ & $1.94$ & $1.94$ & $0.87$ & $1.94$ & $0.87$ \\
$Ste$ & $0.74$ & $2.29$ & $2.29$ & $0.74$ & $2.29$ & $0.74$ \\
$T_{g,0}/T_c$ & $1.85$ & $1.38$ & $1.38$ & $1.85$ & $1.38$ & $1.85$ \\
$\lambda_{\rho}$ & $30.57$ & $35.43$ & $35.43$ & $30.57$ & $35.43$ & $30.57$ \\
$\lambda_{\mu}$ & $3.18$ & $9.34$ & $9.34$ & $3.18$ & $9.34$ & $3.18$ \\
$\lambda_{k}$ & $8.10$ & $2.20$ & $2.20$ & $8.10$ & $2.20$ & $8.10$ \\
$\lambda_{c_p}$ & $4.50$ & $2.11$ & $2.11$ & $4.50$ & $2.11$ & $4.50$ \\
$\lambda_{M}$ & $0.58$ & $3.46$ & $3.46$ & $0.58$ & $3.46$ & $0.58$ \\
\hline
\end{tabular}
}
\label{tab1}
\end{table}

%
\section{Results}
\nsc{
In this study, we compare the evaporation processes of ammonia and n-heptane under different conditions. The results are presented in three parts. First, in section~\ref{sec:sameliq}, we compare the two cases with the same initial liquid volume fraction. Next, in section~\ref{sec:same_en}, we extend this analysis by considering cases where the initial volume fraction of the liquid droplets are chosen to match the available lower heating value (LHV) of the two fuels. Finally, in section~\ref{sec:turb_evap}, we examine the effect of the turbulence intensity on the evaporation process for the different fuels.
}
\subsection{Ammonia and n-heptane evaporation rate at equal liquid volume fraction}\label{sec:sameliq}
\begin{figure}[h!]
\centering
\includegraphics[width = 0.99\textwidth]{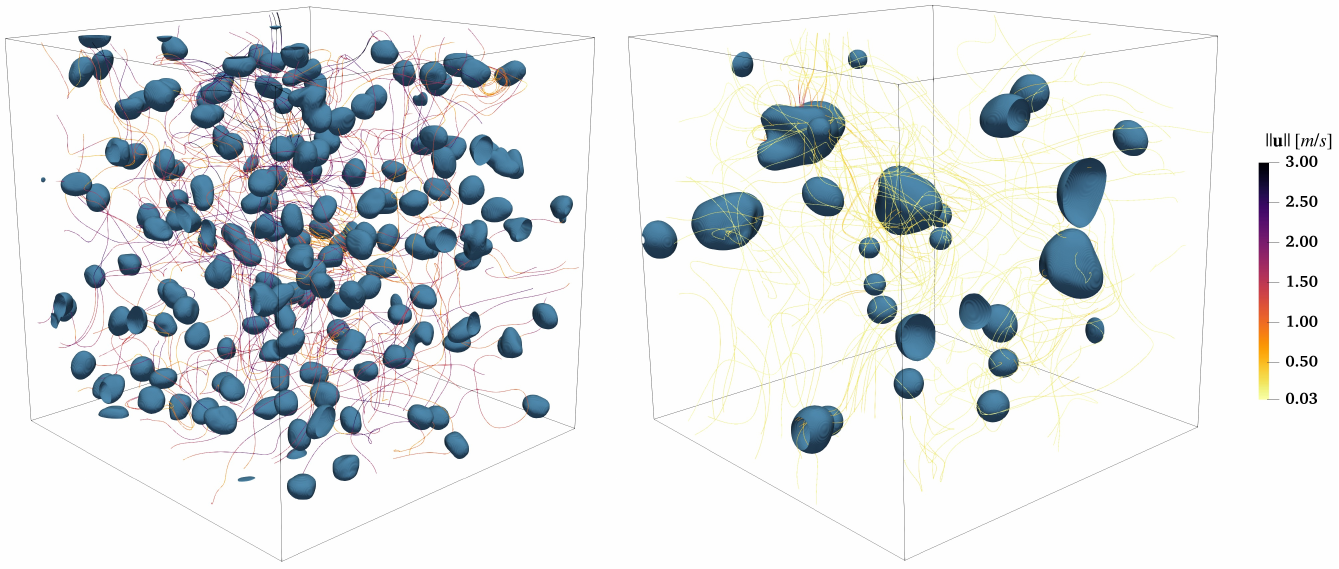}
  \put(-385,180){\small\textbf{(a)}}
  \put(-200,180){\small\textbf{(b)}}
	\caption{Instantaneous streamlines for the simulation Case 1 at time $t = 0.1 \: \text{ms}$ (a) and $t = 5 \: \text{ms}$ (b), colored by the velocity magnitude $\| u \|$. The droplet interface is shown in blue by the $C = 0.5$ isosurface of the phase indicator function.} 
\label{fig:stream_vis}
\end{figure}

\begin{figure}[h!]
\centering
\includegraphics[width = 0.59\textwidth]{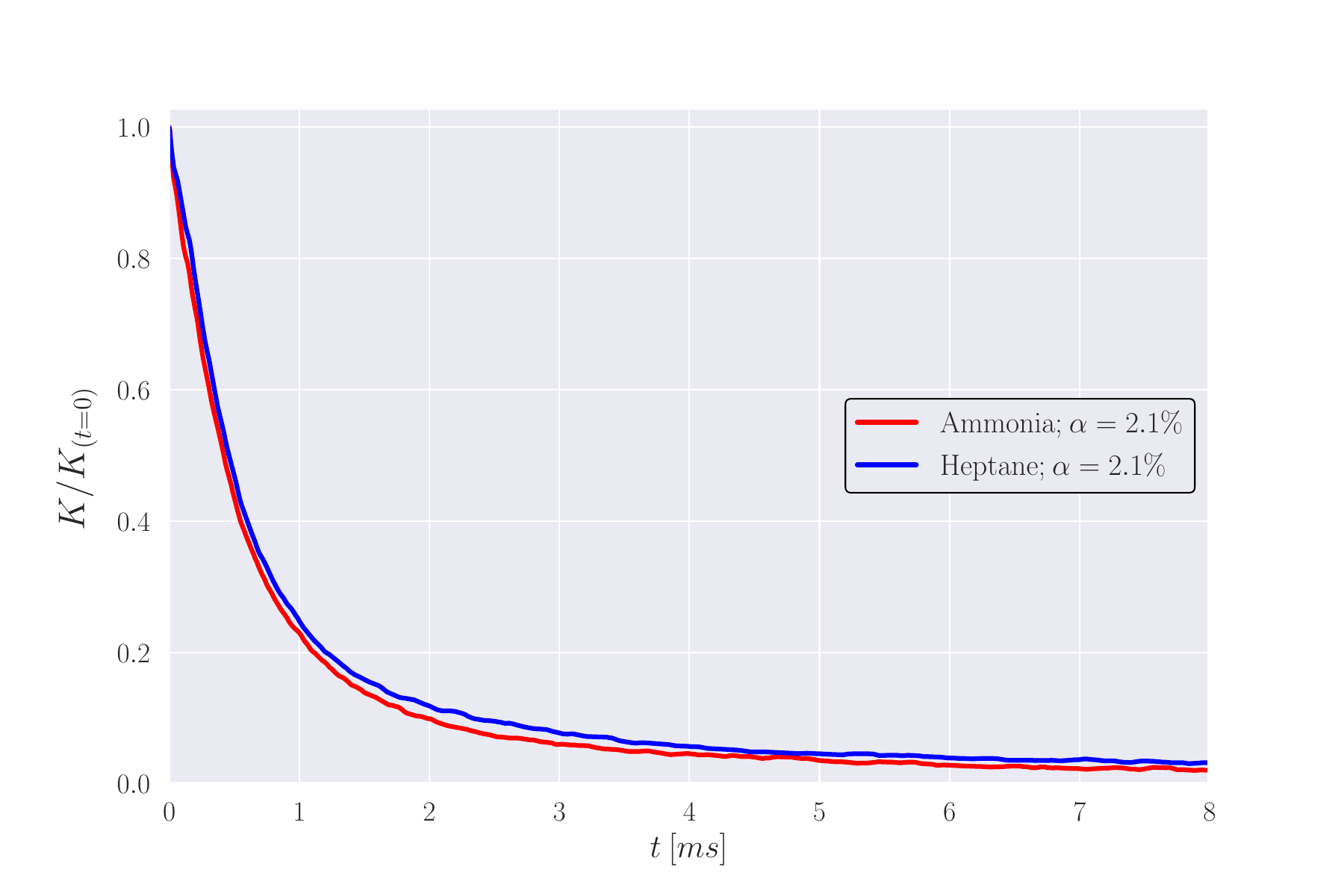}
	\caption{Temporal evolution of the spatially averaged turbulent kinetic energy ($K$), normalized by its initial value, for simulation Case 1 and Case 2.}
\label{fig:tke}
\end{figure}

In this section, we present the DNS results for droplets of two different fuels with the same initial volume fraction, undergoing evaporation, i.e.\ Cases 1 and 2 in table~\ref{tab1}. As the turbulence decays, the droplets deform, break, and coalesce, as shown in Fig.~\ref{fig:stream_vis}. The instantaneous visualization also shows fewer and larger droplets at a later stage, which suggests a predominance of coalescence once the turbulence has decayed. The temporal decay of the spatially averaged turbulent kinetic energy ($K$) is displayed in Fig.~\ref{fig:tke} for Case 1 and Case 2, normalized by the initial value corresponding to the time instance at which the liquid phase is introduced into the system.
%

\begin{figure}[t!]
\centering
\includegraphics[width = 0.99\textwidth]{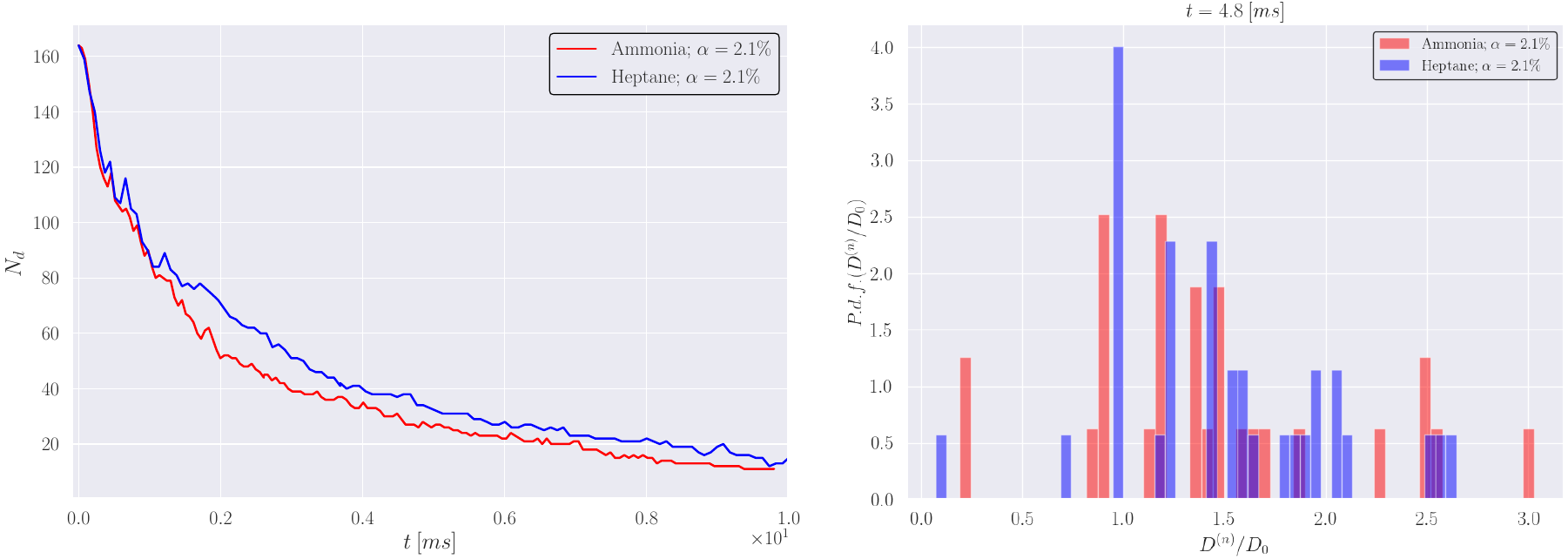}
  \put(-380,140){\small\textbf{(a)}}
  \put(-175,140){\small\textbf{(b)}}

	\caption{(a): temporal evolution of the number of droplets for simulation Case 1 and Case 2. (b): probability density function (PDF) of the normalized droplets diameter at $t = 4.8 \:ms $. } 
\label{fig:num_d_A}
\end{figure}

To provide more quantitative data on the droplets dynamics, we report in Fig.~\ref{fig:num_d_A}(a) the time evolution of the droplet number as they merge and break apart. The decreasing trend confirms the observation from the visualization of the predominance of merging. As turbulence is decaying, the pressure fluctuations are less and less strong to break the droplets.

\begin{figure}[H]
\centering
\includegraphics[width = 0.99\textwidth]{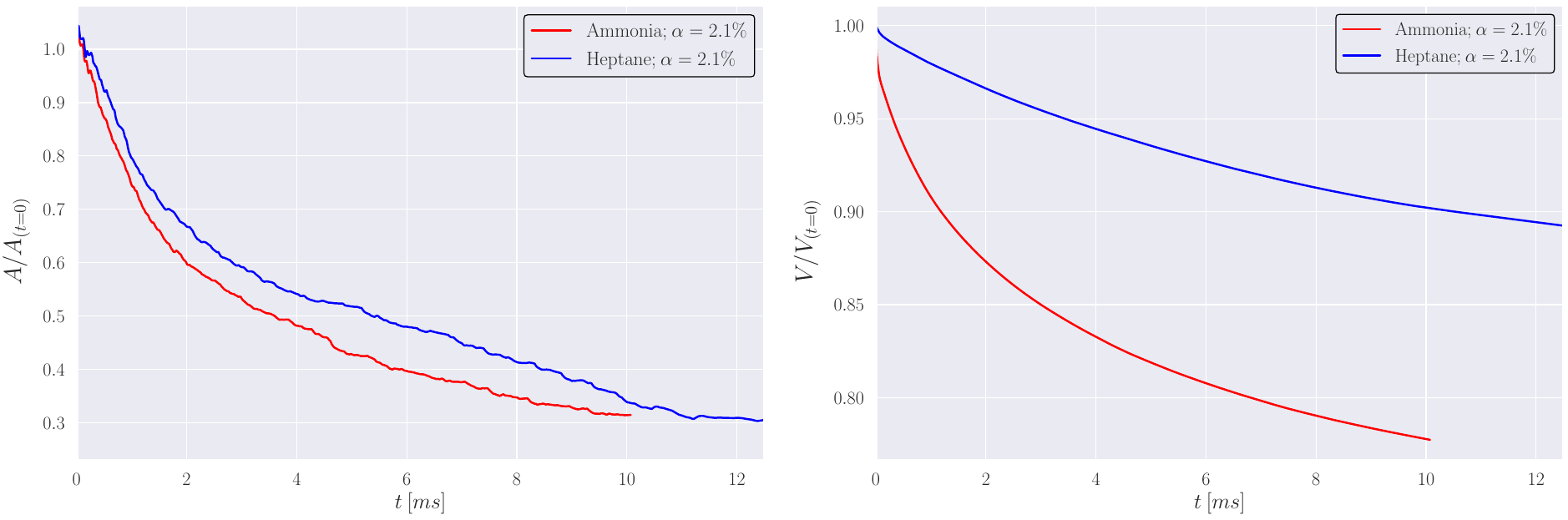}
  \put(-380,140){\small\textbf{(a)}}
  \put(-185,140){\small\textbf{(b)}}
	\caption{(a): Temporal evolution of the normalized total liquid surface area for Case 1 and Case 2. (b): Temporal evolution of the normalized total liquid volume for the same two cases.} 
\label{fig:area_A}
\end{figure}

Fig.~\ref{fig:num_d_A}(b) displays the probability distribution function (PDF) of the droplet diameters at time $t = 4.8 \:ms $, cf.\ the evolution on the panel (a). The data demonstrate that ammonia droplets experience a higher rate of coalescence than n-heptane, a trend that is also confirmed in the temporal evolution of the total surface area, as shown in Fig.~\ref{fig:area_A}(a). This behavior can be attributed to the noticeably higher density ratio ($\lambda_{\rho}$) of n-heptane compared to ammonia. A higher density ratio has been identified by \cite{Dodd_Ferrante_2016} to result in an increased breakup rate and a reduced coalescence rate, thereby leading to a higher total surface area.

While the area evolution provides information on the surface over which evaporation takes place, the liquid-phase volume directly quantifies the evaporation.
The comparison of the temporal evolution of the normalized total droplet volume (Fig.~\ref{fig:area_A}(b)) shows that ammonia undergoes more extensive evaporation than n-heptane throughout the simulation, despite consistently having a smaller total evaporating surface area. This can be primarily attributed to the higher vapor pressure of ammonia compared to n-heptane, as well as the higher diffusivity of vaporized ammonia (corresponding to a lower Schmidt number) than n-heptane.\par

\begin{figure}[t]
\centering
\begin{overpic}[width=0.99\textwidth]{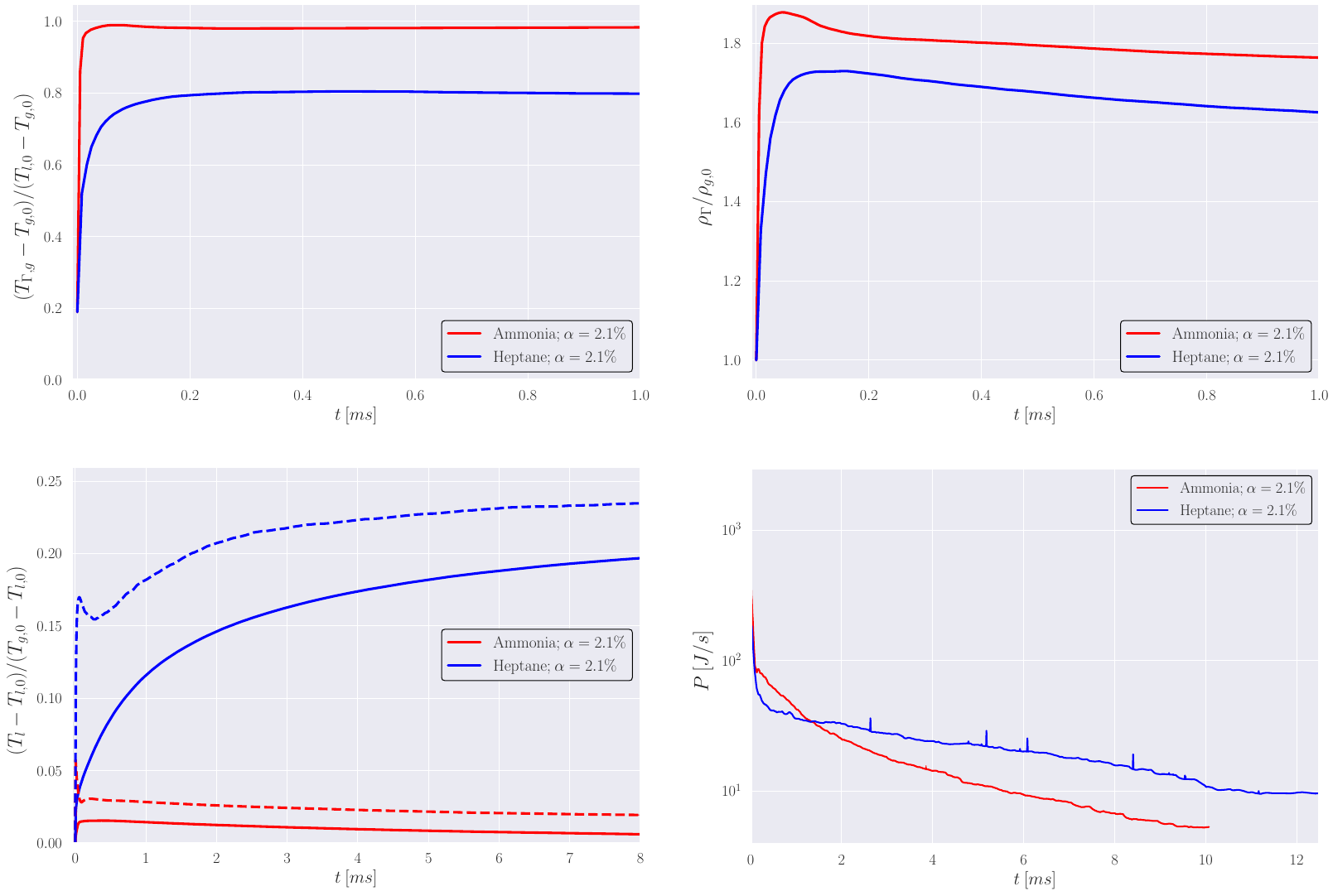}
  \put(0,70){\small\textbf{(a)}}
  \put(50,70){\small\textbf{(b)}}
  \put(0,33){\small\textbf{(c)}}
  \put(50,33){\small\textbf{(d)}}
\end{overpic}
\caption{(a): Temporal evolution of the normalized average interface temperature for Cases 1 and 2. (b): Temporal evolution of the average interface gas density. (c): Temporal evolution of the liquid phase temperature field, averaged over the whole volume of the liquid phase (continuous line) and over the interface between the 2 phases (dashed line). (d): Temporal evolution of the power input, $P = \dot{m}_{t,\Gamma} \: Q_{LHV}$, to the gas phase for Cases 1 and 2. } 
\label{fig:tmp}
\end{figure}
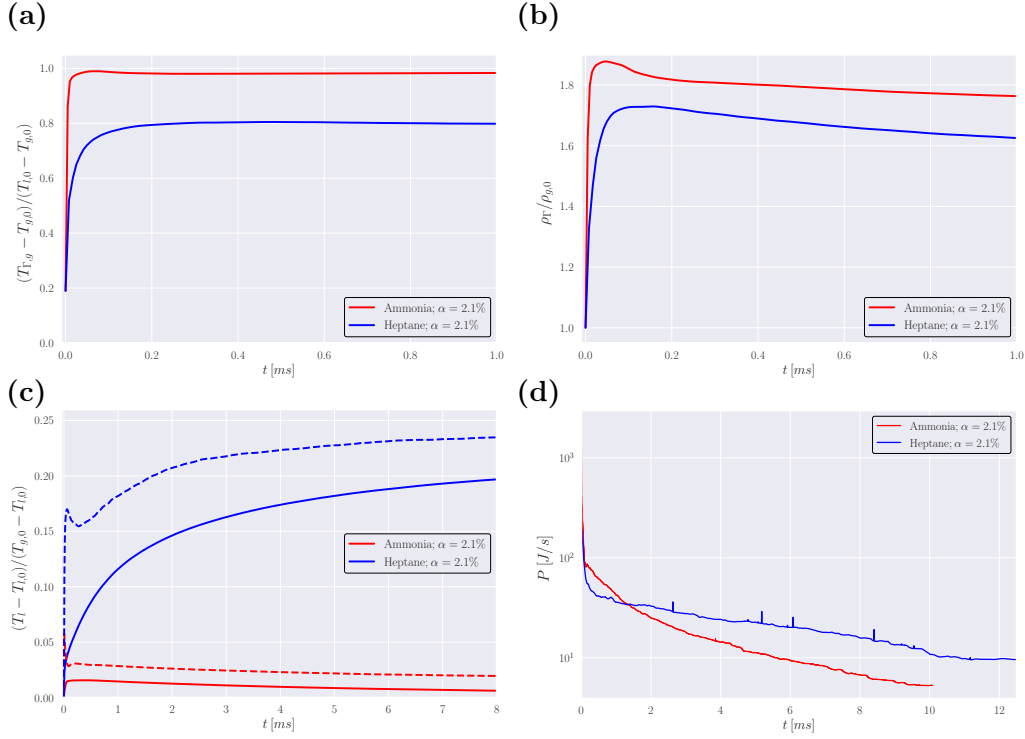

Next, we investigate the averaged properties along the interface and compare the results for the two fuels under consideration in Fig.~\ref{fig:tmp}. Panel (a) presents the normalized gas temperature integrated over the interface. Specifically, it shows $\frac{T_{\Gamma,g} - T_{g,0}}{T_{l,0} - T_{g,0}}$, where $T_{\Gamma,g}$ is the gas temperature integrated over the interface, and $T_{g,0}$ and $T_{l,0}$ are the initial gas and liquid temperatures, respectively.

\begin{figure}[t]
\centering
\begin{overpic}[width=0.99\textwidth]{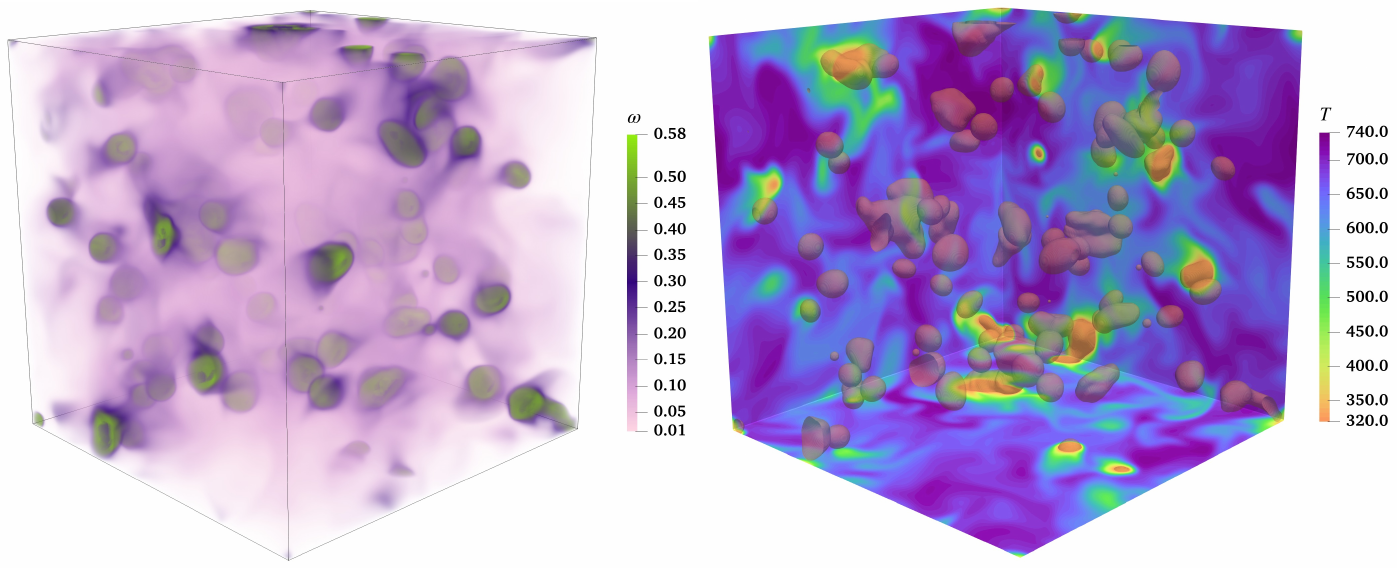}
  \put(50,45){\small\textbf{(b)}}
  \put(0,45){\small\textbf{(a)}}
  \put(95.2,32.2){\fontsize{4}{7}\textbf{[k]}}
\end{overpic}
	\caption{(a): Volume rendering of the vapor mass fraction $\omega$ for simulation Case 1 at $t = 1 \: \text{ms}$. (b): Instantaneous contours of the temperature profile along the domain boundaries, with the translucent surface representing the interface ($C = 0.5$ contour).} 
\label{fig:vap_visu}
\end{figure}

The results show that the temperature of the initially heated gas phase ($T_{\Gamma,g}$) decreases rapidly over time near the interface, reaching saturation conditions. This effect is significantly more pronounced for the ammonia fuel, which quickly reaches the initial temperature of the colder liquid phase. This strong cooling effect is attributed to the higher evaporation rate of the ammonia droplets. A similar trend is observed in Fig.~\ref{fig:vap_visu}(b), where the temperature drop in the vicinity of the droplets is visualized. \par
Fig.~\ref{fig:tmp}(b) illustrates the temporal variation of the density field averaged over the interface. As expected, the case with ammonia fuel exhibits a greater increase in interface density due to the more substantial temperature drop. 

\par
The temporal evolution of the liquid phase temperature is shown in Fig.\ref{fig:tmp}(c), where we report values of the temperature averaged over the entire liquid phase (solid line) and over the interface (dashed line). The results indicate that while the temperature of the n-heptane droplets increases during the evaporation process, the temperature of the ammonia droplets remains close to its initial value due to a pronounced cooling effect. As time progresses, the temperature profiles asymptotically stabilize due to the decreasing evaporation rate, as the domain becomes saturated with the vaporized liquid, as visualized in Fig.~\ref{fig:vap_visu}(a). \par
Finally, Fig.~\ref{fig:tmp}(d) illustrates the temporal evolution of the power input, \sz{defined as $P = \dot{m}_{t,\Gamma} \: Q_{LHV}$, where $\dot{m}_{t,\Gamma}$ denotes the total interfacial mass flux and $Q_{LHV}$ represents the lower heating value, approximately $18$~MJ/kg for ammonia and $44$~MJ/kg for n-heptane}. This quantity reflects the instantaneous energy transfer rate from the liquid to the gas phase due to evaporation. Although this study considers non-reactive conditions, this metric is highly relevant for combustion systems, as it governs the availability of vaporized fuel that contributes to ignition and flame propagation. 
The results demonstrate that, after an initial transient period, Case 2, corresponding to n-heptane evaporation, consistently delivers a higher energy input throughout the evaporation process.

\subsection{Ammonia and n-heptane evaporation rate at equal energy content}\label{sec:same_en}

In this section, we compare Cases 3 and 4, still considering two different liquid fuels but with equal energy content. Here, \emph{``energy content''} refers to the total potential energy available from complete evaporation and combustion of the droplets, i.e., $E_0 = m_{l,0} Q_{\mathrm{LHV}}$, with $m_{l,0}$ denoting the initial liquid mass and $Q_{\mathrm{LHV}}$ the lower heating value. Given the larger values of $Q_{\mathrm{LHV}}$ for n-heptane ($44 \: MJ/Kg$ versus $18 \: MJ/Kg$ for ammonia), the droplet volume fraction is therefore 4.2\% for ammonia and 1.5\% for n-heptane.

\begin{figure}[t]
\centering
\includegraphics[width = 0.99\textwidth]{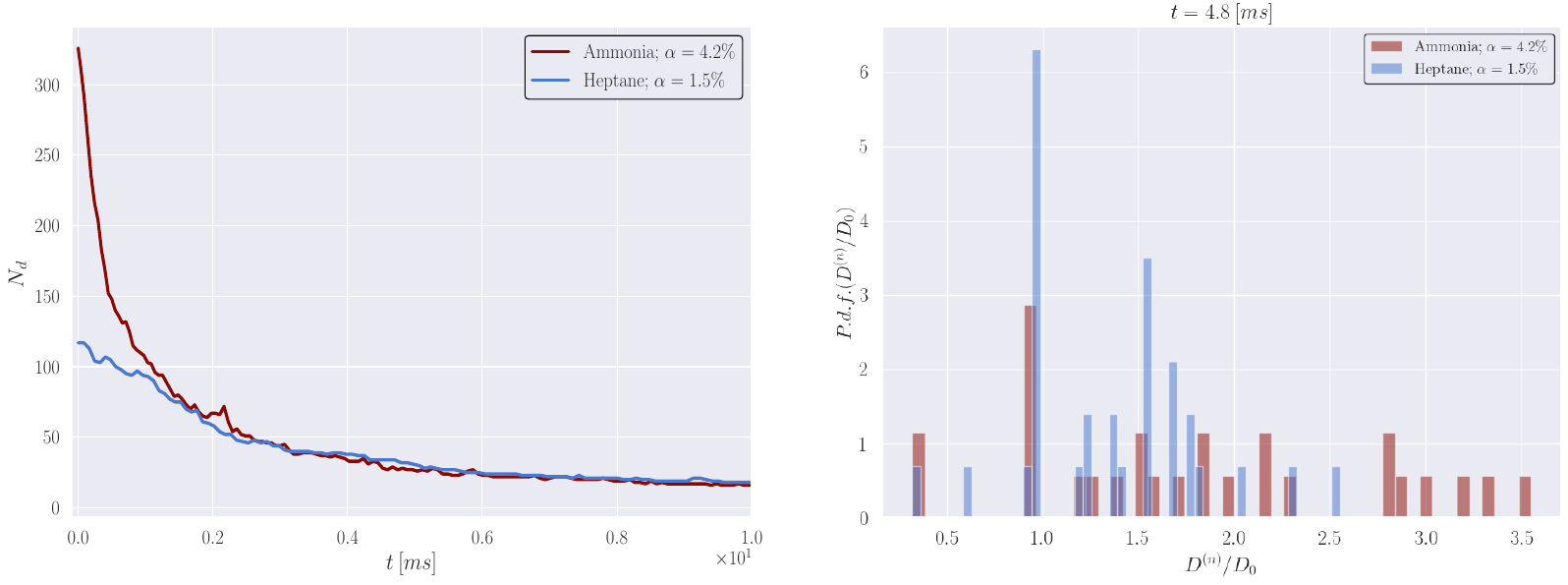}
  \put(-383,150){\small\textbf{(a)}}
  \put(-183,150){\small\textbf{(b)}}
	\caption{(a): Temporal evolution of the droplet count for Case 3 and Case 4. (b): Probability density function (PDF) of the normalized droplet diameters at $t = 4.8 \: \text{ms}$.
} 
\label{fig:num_d_B}
\end{figure}

\begin{figure}[t]
\centering
\begin{overpic}[width=0.99\textwidth]{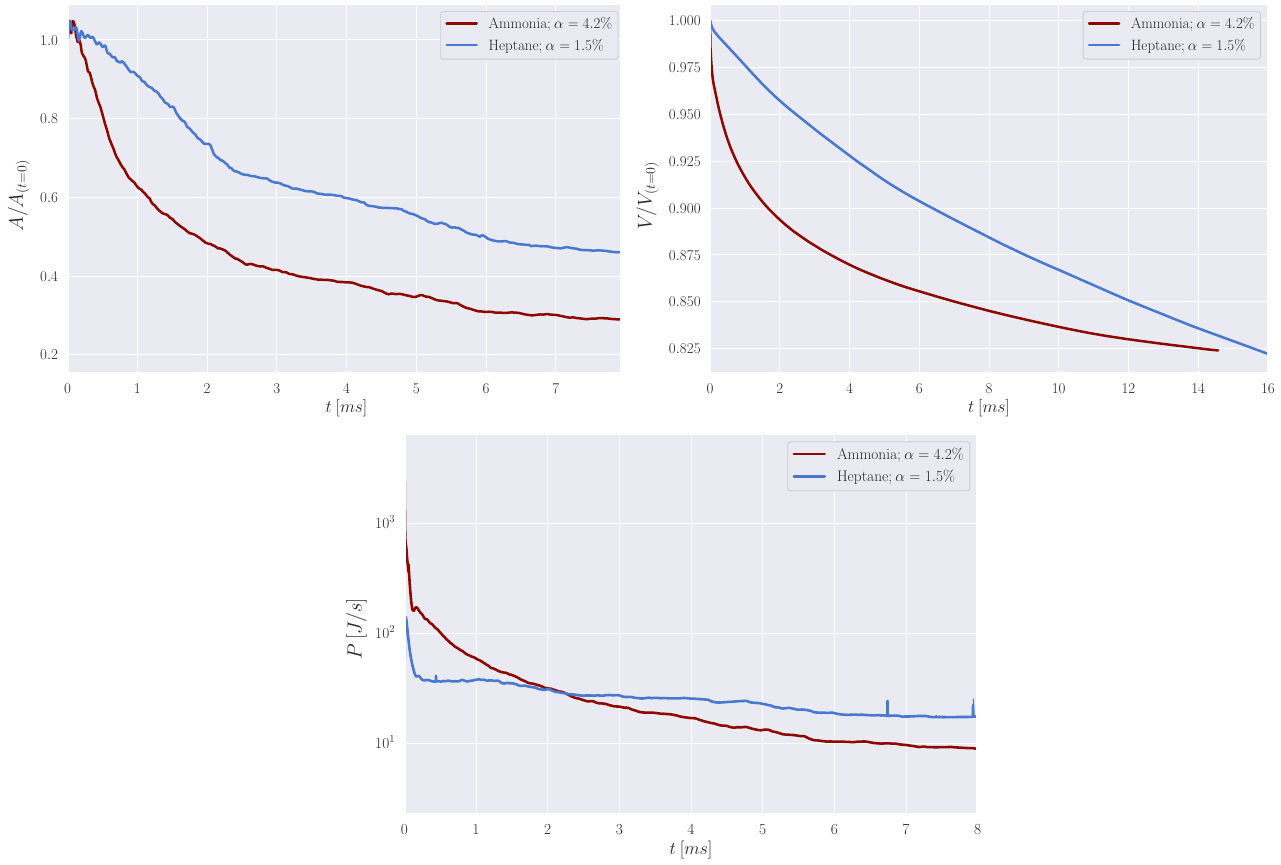}
  \put(0,70){\small\textbf{(a)}}
  \put(50,70){\small\textbf{(b)}}
  \put(27,31){\small\textbf{(c)}}
\end{overpic}
\caption{(a): Temporal evolution of the normalized total surface area for Cases 3 and 4. (b): Temporal evolution of the normalized total liquid volume. (c): Temporal evolution of the power input ($ P= \dot{m}_{t,\Gamma} \: Q_{LHV}$) to the gas phase for Cases 3 and 4.} 
\label{fig:area_B}
\end{figure}

As turbulence evolves and, in this case, decays, the droplets begin to deform, interact with each other, and undergo coalescence and break-ups. Although the two cases start with a different number of initial droplets, the number of droplets eventually converges to similar values, as shown in Fig.~\ref{fig:num_d_B}(a). This seems, therefore, to depend on the turbulence characteristics at the relatively low volume fractions and, later in the simulations, low turbulence intensities considered here.
Consistently, Fig.~\ref{fig:num_d_B}(b) indicates that ammonia droplets undergo a higher rate of coalescence compared to n-heptane in the initial period, resulting in a more marked decrease of the total surface area for the ammonia droplets, as depicted in Fig.~\ref{fig:area_B}(a). 

The temporal evolution of the total volume for the two cases is illustrated in Fig.~\ref{fig:area_B}(b). Initially, the ammonia droplets of Case 4 exhibit a higher evaporation rate, consistent with the findings from the previous section. However, as these droplets continue to coalesce and their total surface area decreases, the evaporation rate slows down. Building upon the earlier analysis, Figure~\ref{fig:area_B}(c) compares the time evolution of the total energy flux rate for both cases. We recall here that this is defined as $P = \dot{m}_{t,\Gamma} \: Q_{LHV}$. The figure indicates that Case 4 initially releases energy at a higher rate than Case 3, consistent with the higher evaporation rate of ammonia, which is explained by the larger interfacial areas due to the larger volume fraction in Case 4. Over time, however, a crossover occurs, and Case 3--with n-heptane droplets, begins to release energy at a higher rate. This is now due to the similar evaporation rate at the late stages of the simulations and the higher energy potential of n-heptane. \par
These findings have important implications for energy efficiency and cost reduction in industrial applications. Understanding the evaporation dynamics and energy release profiles of ammonia and n-heptane droplets enables the optimization of combustion processes. Specifically, the initial high evaporation rate and energy release of ammonia droplets can be advantageous for applications requiring rapid energy delivery. However, the subsequent decrease in evaporation rate due to droplet coalescence highlights the need for system designs that minimize coalescence to maintain efficiency. On the other hand, n-heptane's more consistent evaporation and energy release make it suitable for sustained energy needs.

\subsection{Effect of turbulence intensity on the droplet evaporation rate}\label{sec:turb_evap}

In this final section, we compare cases 3 and 4 discussed in the previous section with cases 5 and 6 of table~\ref{tab1} in order to investigate how turbulence intensity affects the evaporation of the two different fuels under investigation. Specifically, we consider the effect of turbulence with $Re_{\lambda}=80$ and 140, on liquid evaporation at volume fraction $\phi= 1.5\%$ and 4.2\%.

Dodd et al.\ \cite{DODD2021} previously showed that increasing the Reynolds number \( Re_{\lambda} \) enhances evaporation by increasing the extent to which vapor is dispersed; in particular, it is crucial to transport vapor away from the droplet interfaces, which can be efficiently achieved by mixing it by means of the surrounding turbulent eddies in the gas phase. However, their study considered dilute spray conditions with low liquid volume fraction and negligible droplet-droplet interactions. 
In contrast, we consider here a configuration with a higher liquid volume fraction, where extensive droplet-droplet interactions occur. This allows us to examine the effect of turbulence intensity in a denser spray environment, more relevant to spray combustion applications, where droplet interactions may play a significant role in the evaporation process. \par
\begin{figure}[t]
\centering
\includegraphics[width = 0.99\textwidth]{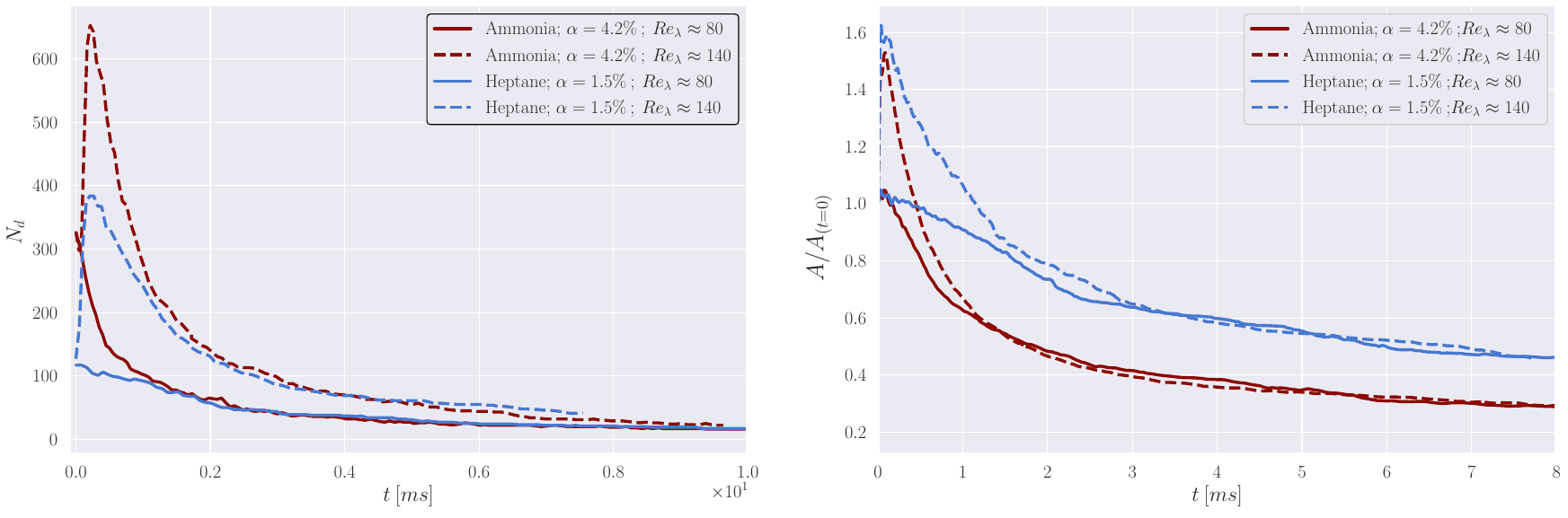}
  \put(-383,140){\small\textbf{(a)}}
  \put(-183,140){\small\textbf{(b)}}
	\caption{(a): Temporal evolution of the number of droplets for different fuels and different turbulence intensities (see Cases 3-6 in Table \ref{tab1} ). (b): Temporal evolution of the normalized total surface area for Cases 3-6. }
\label{fig:num_d_C}
\end{figure}
Fig.~\ref{fig:num_d_C} shows the temporal evolution of the number of droplets and of the normalized surface area. For all cases, with different liquid volume fractions and different $Re_{\lambda}$, in spite of the different initial transient behavior, the number of droplets finally converges to similar values. As concerns the initial transient, we can see an increase in the number of droplets, so a significant breakup, for the highest turbulence intensities (dashed line in the figure). It should also be noted that consistent with the results presented in the previous section, the normalized total surface area is higher in cases 3 and 5, which represent n-heptane droplets, compared to the cases with ammonia droplets. This difference is attributed to the higher coalescence rate observed in the cases with ammonia fuel, driven by their higher initial volume fraction.

%
\begin{figure}[t]
\centering
\includegraphics[width = 0.99\textwidth]{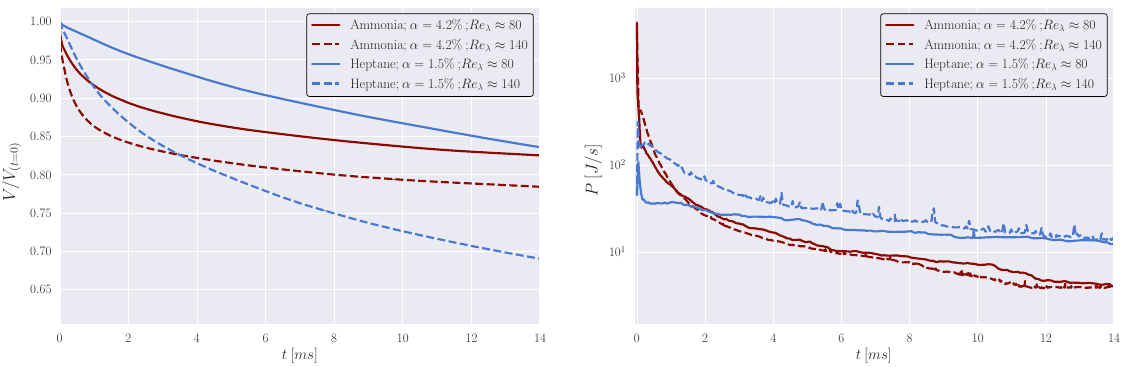}
  \put(-383,140){\small\textbf{(a)}}
  \put(-183,140){\small\textbf{(b)}}
	\caption{(a): Temporal evolution of the normalized total volume for cases 3-6. (b): Temporal evolution of the power input ($ P= \dot{m}_{t,\Gamma} \: Q_{LHV}$) to the gas phase for cases 3-6.} 
\label{fig:ene_C}
\end{figure}

The observations on the surface area and number of droplets provide background information to examine the temporal evolution of the total volume of droplets depicted in Fig.~\ref{fig:ene_C}(a).
The data confirm that an increase in turbulence intensity generally leads to a higher evaporation rate. Consequently, the cases with higher turbulence intensities (indicated by dashed lines in Fig.~\ref{fig:ene_C}(a)) exhibit a more pronounced total volume reduction. However, the results also indicate that increasing the Reynolds number \( Re_{\lambda} \) influences the evaporation behavior of different fuels in distinct ways. In the case of n-heptane droplets (cases 3 and 5), an increase in turbulence intensity results in a significant and sustained enhancement of the evaporation process throughout the entire simulation. Case 6, characterized by higher turbulence intensity ($\phi = 4.2\%$) and ammonia as the working fuel, also exhibits faster evaporation during the initial stages. This behavior is attributed to the increased total surface area resulting from droplet breakups, as shown in Figure \ref{fig:num_d_C}.

The dynamics appear to be different for ammonia droplets (cases 4 and 6 in table \ref{tab1}). Here, we observe that increasing the turbulence intensity initially accelerates the evaporation process, which again we can relate to droplet breakup; however, after approximately $t \approx  2 [ms]$, the evaporation rate tends to same limit (see slope of the curves) at both low and high turbulence intensity. At these late stages, the ammonia has diffused in the gas phases, reaching an almost uniform distribution, which, therefore, harms the additional mixing that turbulence would provide. Note that this is not the case for n-heptane: the mass flux is indeed slightly higher at $Re_{\lambda}=140$ also at the late simulation times due to the additional mixing provided by the turbulence when vapor diffusion requires longer times.
\par
The last observation can be better understood by examining Fig.~\ref{fig:ene_C}(b), which shows the temporal evolution of the rate of energy supply (power input) from the liquid to the gas phase for the different cases, which is proportional to mass flux when considering the same fuel. As discussed above, the evaporation rate is higher at the highest turbulence intensity for n-heptane until the late simulation stages, when the vapor is eventually well mixed in the gas phase. For ammonia, only the initial transient, characterized by more numerous and smaller droplets, displays a higher evaporation rate when turbulence is strongest. Soon the evaporation rate is similar, even slightly larger for the smallest turbulence. This last effect is attributed to the larger droplet size, as a consequence of the lower early evaporation at low turbulence intensity.

Due to the very different heating values of the two different fuels, ammonia, and n-heptane, it is relevant to study this process in further detail. The power input from the liquid to the gas phase is the controlling parameter in spray combustion applications where the initial ignition process and the later flame propagation rate are controlled by the amount of fuel available in the gas phase. Except for the very initial phase of the evaporation process, characterized by a breakup in our computational setup, the liquid ammonia droplets provide fuel to the gas phase at a lower rate compared to n-heptane. This result is due to a combination of vapor saturation in the gas phase, taking place in the ammonia case after the initial rapid evaporation, and of the significantly lower specific energy content of the ammonia compared to n-heptane ($18$ vs. $44$~MJ/Kg). Interestingly hence, while the transient rate of energy supply to the gas phase for ammonia is basically independent of the turbulence intensity, this is not the case for n-heptane that shows a significant $Re_{\lambda}$ dependence in the crucial (for ignition, combustion) transient phase. To investigate whether and when turbulence would affect also the evaporation of the faster-diffusing ammonia, one would therefore need to consider higher turbulence intensities, higher than those currently affordable by our numerical simulations.
\par
\begin{figure}[t]
\centering
\includegraphics[width = 0.99\textwidth]{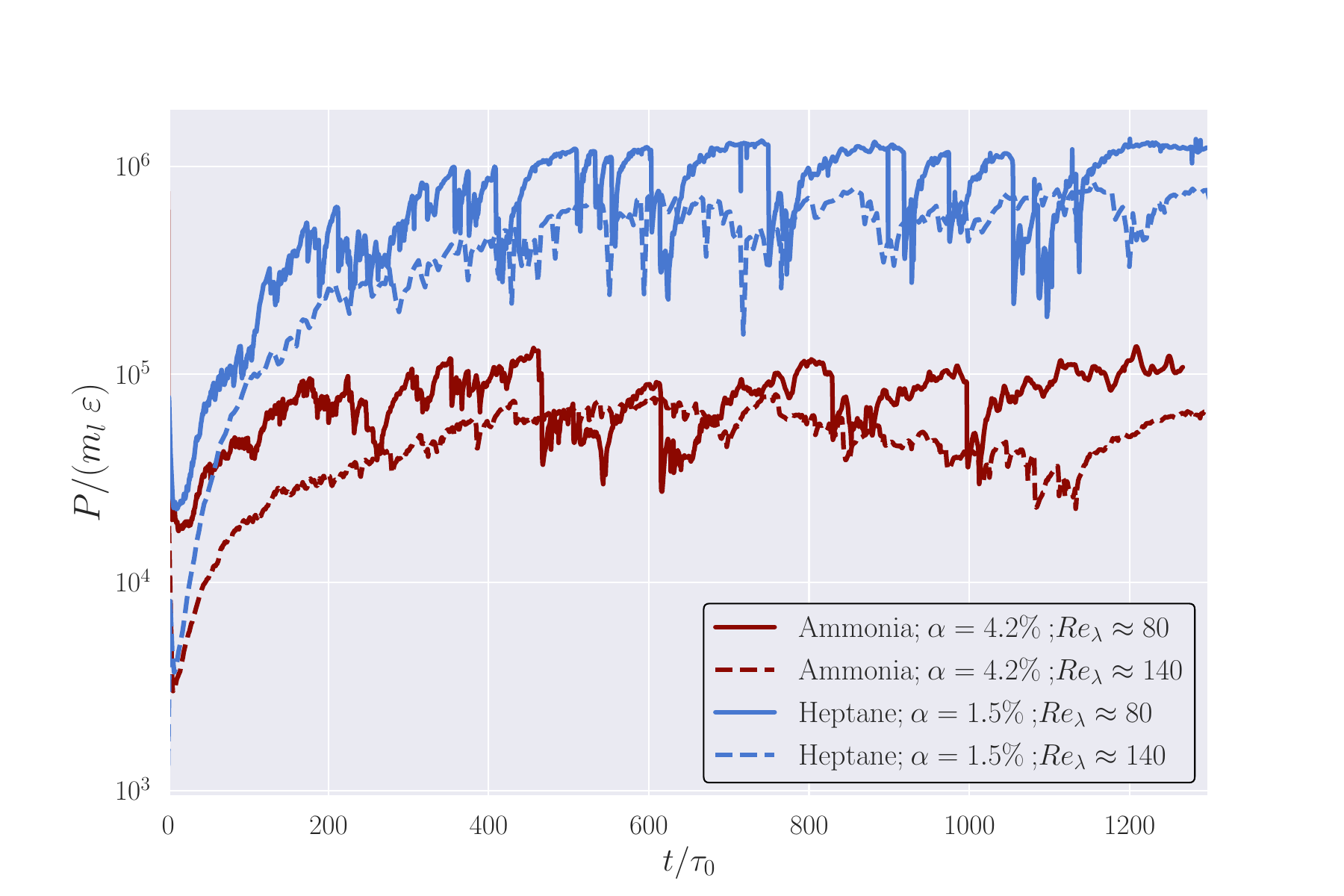}
	\caption{Temporal evolution of the normalized power input to the gas phase. The instantaneous rate of energy supply (power input) from the liquid to the gas phase is normalized by the instantaneous liquid mass ($m_l$) and the turbulent energy dissipation ($\varepsilon$). The time is scaled by the initial turbulent turnover time ($\tau_0$).} 
\label{fig:ene_C_nondim}
\end{figure}
The fundamental characteristics of the transient evaporation process is conveniently illustrated in Fig.~\ref{fig:ene_C_nondim} by the normalized instantaneous rate of energy supply (power input) from the liquid to the gas phase. The normalization is implemented by the instantaneous liquid mass ($m_l$) multiplied by the turbulent energy dissipation ($\varepsilon$) while time is expressed in units of the initial turbulent turnover time $\tau_0 \approx K_0/\varepsilon_0$, where $K_0$ is the initial averaged turbulent kinetic energy and $\varepsilon_0$ is the initial averaged turbulence dissipation rate. The figure seems to suggest that, except for stochastic short-lived oscillations likely caused by droplet-droplet interactions, i.e., coalescence and breakup events, the evaporation rate and interphase energy-transfer behavior can be represented by a (normalized) scaling law that is independent of the turbulent intensity for the two different fuels. This finding, if confirmed by additional simulations at different conditions, constitutes potentially important information for accurate low-order models of spray evaporation and combustion.

\section{Conclusions }
In this study, we have used direct numerical simulations (DNS) to explore the evaporation process of ammonia and n-heptane droplets in a \agr{non-reacting} turbulent-flow environment, varying both the liquid volume fractions and turbulence intensities. When comparing \agr{the evaporation rate of} ammonia and n-heptane droplets with the same initial liquid volume fraction, ammonia initially exhibits a higher evaporation rate due to its higher volatility. However, as the ammonia droplets experienced coalescence more frequently, the total \agr{liquid-gas interface} surface area decreased over time, leading to a late decay in the ammonia evaporation rate. In contrast, n-heptane droplets maintain a more constant evaporation rate profile, with a lower degree of coalescence and a slow but steady rate of evaporation. \par
\agr{We have also compared cases configured to match the total amount of chemical energy (lower heating value - LHV) present within the computational domain, resulting in 4.2\% and 1.5\% volume fractions for ammonia and n-heptane given the different
lower energy content of the two. The data indicate that the ammonia fuel droplets initially evaporate faster and energy is transferred to the gas phase at a higher rate compared to the case of n-heptane droplets. However, as ammonia droplets coalesced more frequently, its evaporation rate also slows down in this comparison, leading to a crossover time, after which n-heptane droplets eventually transfer energy to the gas phase at a higher rate. This result suggests that, in the context of a fuel switch between conventional diesel and ammonia, the liquid-jet injection and break-up patterns, leading to atomization in fuel injectors, need to be carefully adapted to take into account the different evaporation-rate profiles of these two fuels.} \par
Simulations on the impact of the turbulence intensity on the evaporation rate reveal that an increased Reynolds number $Re_{\lambda}$ significantly enhances the evaporation rate for n-heptane droplets throughout the simulation. Conversely, in the case of ammonia droplets, higher turbulence intensity initially accelerates the evaporation rate owing to droplet breakup; nevertheless, the evaporation rates at both low and high turbulence intensities converge to similar values after the initial transient with fast evaporation. This suggests that while the presence of stronger turbulent motions can enhance the evaporation rate for both fuels, the effect is more sustained in time for n-heptane, whereas the ammonia evaporation rate reaches a saturation point due to faster vapor diffusion in the gas phase. \par
We believe that the data generated from this study provide valuable insights for improving predictive models of droplet evaporation in turbulent flows, contributing to more effective designs of spray combustion systems at high pressures, and hope they can trigger new investigations on the effect of turbulent mixing in shear flows and sprays, as examples of open questions.

\bibliography{bibfile}

\end{document}